\documentclass[lettersize,journal]{IEEEtran}

\usepackage{amsmath,amssymb,amsfonts}
\usepackage{cuted}      
\usepackage{stfloats}   
\usepackage{graphicx}
\usepackage[caption=false,font=normalsize,labelfont=sf,textfont=sf]{subfig}

\usepackage{algorithm,algorithmic}
\usepackage{array}
\usepackage{textcomp}
\usepackage{url}
\usepackage{verbatim}
\usepackage{cite}
\usepackage[mathscr]{euscript}
\usepackage{color}
\usepackage{soul}
\usepackage{needspace}
\usepackage{amsmath}
\usepackage{wasysym}

\usepackage{cleveref}

\definecolor{LightBlue}{rgb}{0.75,0.936,1.00}
\sethlcolor{LightBlue}
\definecolor{Yellow}{rgb}{1.00, 1.00, 0.2}
\sethlcolor{Yellow}
\definecolor{LightCyan}{rgb}{0.88,1,1}

\begin{document}

\title{On the Performance of RIS-Empowered Transmitters with Receive Combining Techniques}
\author{\thanks{Manuscript received Month XX, 2026. The associate editor coordinating the review of this paper and approving it for publication was XXXX 2026.}\thanks{Muhammed Faruk Karabay and Oğuz Kucur are with the Department of Electronics Engineering, Gebze Technical University, Turkey (Email: \{faruk.karabay, okucur\}\,@gtu.edu.tr)}\IEEEauthorblockN{Muhammed Faruk Karabay, Oğuz Kucur and Ertugrul Basar,~\IEEEmembership{Fellow,~IEEE}} \thanks{Ertugrul Basar is with the Tampere Wireless Research Center, Department
of Electrical Engineering, Tampere University, 33720 Tampere, Finland (Email: ertugrul.basar@tuni.fi)}}
\vspace{-5mm}



\markboth{TO BE SUBMITTED TO IEEE JOURNALS, VOL. XX, NO. XX, XXXX 2026}
{Shell \MakeLowercase{\textit{et al.}}: Bare Demo of IEEEtran.cls for IEEE Journals}

\maketitle
\begin{abstract}
Reconfigurable intelligent surface (RIS)-empowered transmitters offer a promising basis for hardware-efficient multi-antenna architectures in next-generation wireless networks. In particular, combining these transmitters with low-complexity reception techniques such as receive antenna selection (RAS) can further simplify the receiver architecture. In this paper, we investigate the impact of practical phase-shift constraint on RIS-empowered multi-antenna systems employing maximal ratio combining (MRC) and RAS. Since each RIS element can apply only a single phase-shift at a given time, the system performance degrades substantially when multiple receive antennas are active simultaneously. To quantify this effect, closed-form outage probability expressions are derived and asymptotic analyses are carried out over Rician fading channels. We show that the practical phase-shift constraint of RIS-based transmitters can degrade the diversity order of MRC and in addition, RAS outperforms MRC in the examined system models, contrary to the conventional performance ordering of these techniques.
\end{abstract}


\section{Introduction}

Future wireless communication systems impose increasingly demanding requirements in terms of data rate, reliability, and energy efficiency \cite{basar2019wireless}. Conventionally, the wireless propagation environment has been considered as uncontrollable, while performance improvements have mainly relied on active transceiver architectures, advanced signal processing and cooperative relaying. As effective as these solutions are, when it comes to improving performance, they may increase hardware complexity and power consumption \cite{basar2019wireless,intro1,intro2}. Therefore, recent research has shifted toward intelligent propagation environments, where the channel can be partially shaped to improve communication performance. Reconfigurable intelligent surfaces (RISs) have attracted significant attention in this regard, as they can manipulate the wireless channel through low-cost and nearly passive reflecting elements. With the ever-increasing demand for higher data rates, the need to improve coverage due to the use of high frequencies \cite{BS}, and its ability to further increase reliability when combined with multiple antennas at the receiver side \cite{hassain}, RIS has a unique place in fifth-generation (5G) and beyond systems.

There are many studies that focus on dual-hop RIS-assisted systems. However, one of the main drawbacks of dual hop RIS systems is the double fading effect \cite{transmit1}. This challenge can be overcome by various structures, such as active RIS \cite{transmit1,double}, RIS-integrated multiple input multiple output (MIMO) systems \cite{ferkan,benimki,star}, simultaneously transmitting and reflecting RIS (STAR-RIS) \cite{star,papazSTAR} and proper RIS positioning \cite{pos1,pos2}. It has been shown both theoretically \cite{posTheo} and experimentally \cite{pos1} that deploying an RIS near the transmitter or the receiver improves performance by alleviating the cascaded path-loss attenuation. Beyond deployment optimization, another promising direction is to exploit the RIS itself as a part of the transmission mechanism. Using an RIS as a transmitter has many use cases, such as cost and energy efficient transmission through modulation \cite{basar2019wireless,transmit1}, index modulation \cite{transmit2}, transmit diversity schemes, namely the Alamouti technique \cite{ala1}. It was also shown in \cite{pratik} that using an RIS as a transmitter enables robust and cost-effective communication. In this regard, this approach provides an alternative perspective to conventional dual hop RIS-assisted transmission.

In parallel, RIS-assisted MIMO systems have also been widely investigated due to their potential to improve spectral efficiency, reliability, and diversity performance \cite{BS,hassain,ferkan,benimki,papazSTAR,star,mimolu1,mimolu2}. In many works, the RIS phase-shifts are generally designed through optimization frameworks together with transmit beamforming, power allocation, or active amplification coefficients, with the objective of maximizing capacity, spectral efficiency or achievable sum-rate \cite{BS,papazSTAR,star,mimolu1,mimolu2}. Complementary to these optimization-oriented RIS-assisted MIMO studies, several works have investigated RIS-assisted systems with maximal ratio combining (MRC), focusing on theoretical performance analysis rather than joint beamforming optimization with the purpose of analyzing outage probability (OP), error performance and high signal-to-noise ratio (SNR) asymptotic behavior. 

In theoretical performance analysis of RIS-assisted MIMO systems, one of the most critical points is the definition of phase-shift configuration of the RIS. While the authors in \cite{star} consider random phase-shifts at the RIS, authors of \cite{hassain} and \cite{ferkan} assume perfect phase-shifts. Incorporating practical imperfections into the analysis of RIS-assisted systems is non-trivial and the problem becomes even more challenging in RIS-assisted multi-antenna systems. In fact, to the best of our knowledge, only \cite{benimki} and \cite{hwi1} provide theoretical performance analyses of practical imperfections in RIS-assisted multi-antenna systems, considering phase errors and hardware impairments, respectively. Furthermore, another overlooked practical imperfection of RIS-assisted multi-antenna systems is what we refer to as \textit{phase dependency issue} in combining techniques such as MRC. In RIS architectures, each reflecting element applies a single phase-shift at a given time \cite{basar2019wireless}. Therefore, using multi-antenna techniques such as MRC in RIS-assisted systems limits the full exploitation of MRC, since RIS cannot simultaneously compensate the phases of different MRC branches. The relevance of this issue is further motivated by the potential of RIS-based transmitters for low-hardware-complexity Internet of Things (IoT) communication \cite{iot}. Understanding its impact under simple RIS phase-alignment policies is therefore important for evaluating receiver designs such as MRC and receive antenna selection (RAS) in these architectures. To the best of our knowledge, no existing work analytically investigates the impact of this issue on RIS-assisted combining systems.

The main objectives and contributions of this work can be summarized as follows:

\begin{itemize}
    \item We consider a system where an RIS is used as a transmitter. Using such a structure in practice requires the direct link not to be completely blocked. Therefore, we analyze the system by considering a strong channel between the transmitter and the receiver, meaning that the system is examined under the Rician fading channel assumption for more generalized analysis. 
    \item The system is analyzed using two multiple-antenna techniques, namely MRC and selection combining, also referred to as RAS in this work. Closed form OP expressions are derived in the form of multivariate Fox-H function and diversity order (DO) of both systems are obtained via asymptotic analyses,  which provide insight into the high SNR behavior and the relative performance of the two multi-antenna techniques.
    
    \item Performance analyses in the literature do not consider the \textit{phase dependency issue} and assume either phase-free or randomly phased channels when employing MRC \cite{hassain,ferkan,star}. We show that the practical phase-shift constraint of RIS severely degrades the performance gain of MRC and shifts the performance-gain balance. We further show that RAS performs considerably better than MRC when used in the examined RIS-assisted systems, which defies the optimality of MRC \cite{unified}.
     
\end{itemize}

\begin{figure}[t]
  \centering
  \includegraphics[width=0.95\columnwidth]{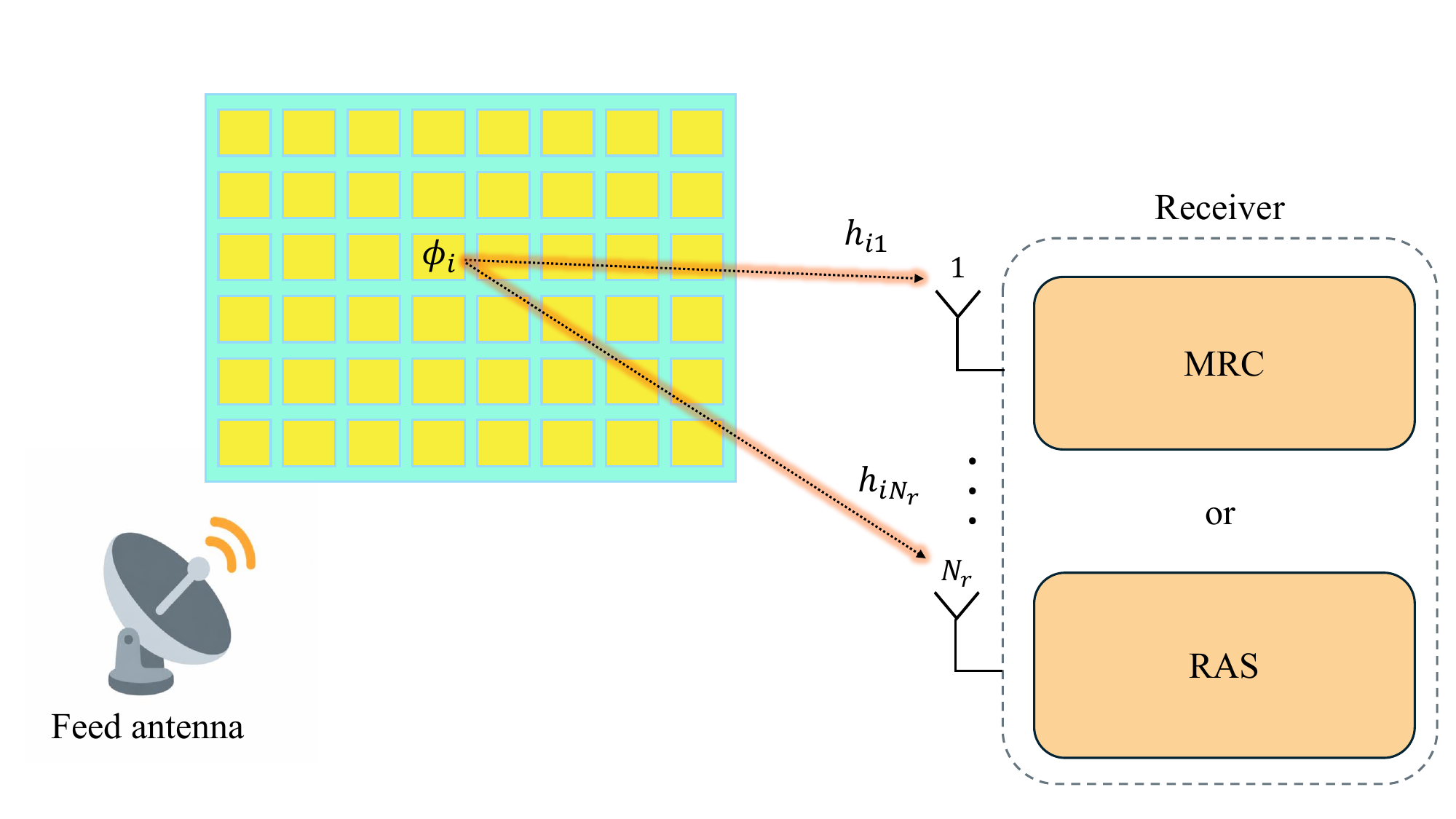}
  \caption{RIS-assisted multi-antenna system with RAS and MRC techniques.}
  \label{fig:sis1}
\end{figure}

The remainder of this paper is organized as follows. Section II introduces the system model. Section III presents the MRC and RAS analyses. In Section IV, we carry out asymptotic analyses and obtain DO expressions of both techniques. Section V shows the numerical results and we conclude the paper with Section VI.

\section{System Model}

We consider a system in which an RIS is used as a transmitter \cite{ala1,basar2019wireless}, and the receiver is equipped with $N_r$ antennas as depicted in Fig. 1. Let $N$ and $\zeta$ represent the number of RIS elements and path-loss coefficient, respectively. $s$ denotes the transmitted signal with energy $E_S$. $\zeta$ is assumed to be the same for all channels. Also, let $h_{ik}=|h_{ik}|e^{j\theta_{ik}}$ denote the channel coefficient between the $i$th RIS element and $k$th receive antenna. The additive Gaussian noise sample at the $k$th receive antenna is denoted by $n_k\sim \mathcal{CN}(0,N_0)$. In practice, the RIS is able to do single phase-shift through each surface in single time frame namely as $\phi_i$. So, the received signal at $k$th receive antenna can be expressed as

\begin{equation}
\begin{aligned}
    r_k &= \sqrt{\zeta} \sum_{i=1}^{N} h_{ik}\, e^{-j\phi_i} s + n_k \\
        &= \sqrt{\zeta}\,\, \underbrace{\sum_{i=1}^{N} |h_{ik}|\, e^{j(\theta_{ik}-\phi_i)}}_{X_k}\,\, s + n_k.
\end{aligned}
\label{eq:ilk}
\end{equation}

Here, $X_k$ denotes effective channel coefficient observed at the $k$th receive antenna. All channel coefficients are assumed to be independent and identically distributed (i.i.d.), and the channel amplitudes $|h_{ik}|$ follow Rician distribution with unit power ($\Omega=1$). So, SNR of $k$th antenna is expressed as $\gamma_k = \zeta \, \bar{\gamma} |X_k|^2$, where $\bar{\gamma}=E_S/N_0$.

\section{Performance Analysis}
In this section, we  analytically investigate the system model using MRC and then RAS.
\subsection{MRC Analysis}
 Based on the system model described above, we first analyze the performance of the considered system under MRC. For MRC, all radio-frequency (RF) chains must remain active throughout data reception. However, each RIS element applies a single phase shift at a given time, resulting in a common phase configuration shared by all receive antennas. Jointly optimizing the RIS phases to maximize the post-combining SNR generally leads to a nonconvex optimization problem, even under continuous phase control \cite{complex}. Moreover, the dependence of the optimized phase configuration on the channels of all receive antennas complicates the statistical characterization of the resulting SNR. Instead, we select one receive antenna as the reference one and align the RIS phases accordingly, thereby maximizing the SNR at the selected antenna. Although this approach may not yield the maximum possible post-combining SNR, it provides an explicit phase-control rule and enables analytically tractable statistical characterization. Owing to the statistical symmetry of the receive-antenna channels before RIS phase configuration, we select the first receive antenna as the fixed reference without loss of generality. The effective channel coefficients after applying the continuous phase shifts (i.e., $\phi_i=\theta_{i1}$) are given by 

\begin{equation}
X_k=
\begin{cases}
    \sum_{i=1}^{N}\,|h_{i1}|, & k=1 \\
    \sum_{i=1}^{N}\,|h_{ik}|\,e^{j(\theta_{ik}-\theta_{i1})}, & k=2,\cdots,N_r.
\end{cases}
\label{eq:Xk}
\end{equation}

It is clear that, due to the \textit{phase dependency issue}, the RIS phases are fully aligned with respect to the first receive antenna, while the remaining receive antennas experience residual phase-shifts. We first characterize the equivalent channel observed at the $1$st receive antenna. Since $X_1$ is equal to sum of Rician variables, deriving statistics of $X_1^2$ is analytically challenging and does not lead to efficiently computable closed-form expressions. Therefore, we proceed with a highly accurate and tractable approximation, namely the Gamma-Gamma $(\Gamma\Gamma)$ approximation \cite{Gamma}. The distribution of $\gamma_1$ is approximated by a Gamma-Gamma random variable $Y\sim \Gamma\Gamma(a,d,p)$, where $a$, $d$, and $p$ are the distribution parameters. The probability density function (PDF) and the moments of $Y$ are respectively expressed as

\begin{equation}
f_Y(y) = 
\frac{p}{a^d \Gamma\left(\frac{d}{p}\right)}
y^{d-1}
\exp\left[-\left(\frac{y}{a}\right)^p\right],
\qquad y \geq 0.
\label{eq:fy}
\end{equation}
\begin{equation}
M_r = \mathbb{E}\left[Y^r\right]
= a^r \frac{\Gamma\left(q+\frac{r}{p}\right)}{\Gamma(q)},
\label{eq:mr}
\end{equation}where $q=d/p$ and $\Gamma(\cdot)$ is Gamma function \cite{gradshteyn2007table}, $\mathbb{E}[\cdot]$ is expectation operator. In this context, the approximation is obtained by matching the first three moments of $Y$ with those of $\gamma_1$. Let us denote $S_N=\sum_{i=1}^{N}|h_{i1}|=\sum_{i=1}^{N}\rho_{i}$ and $m_r=\mathbb{E}[\rho_{i}^r]$. Moments of Rician amplitudes $\rho_{i}$ have well-known formula and are omitted here for brevity. To obtain the $r$th moment of $S_N$ efficiently, we derive a recursive expression, whose derivation is provided in Appendix~\ref{app:ilkder}. It is given by
\begin{equation}
    \mathbb{E}[S_N^r]=\sum_{j=0}^{r}\binom{r}{j}\mathbb{E}[S_{N-1}^{r-j}]\,m_j
\end{equation}
Then, moments of $Y$ can be matched with $M_r=(\zeta \bar{\gamma})^{r} \mathbb{E}[S_N^{\,2r}]$ and parameters $a,p,q$ of $Y$ can be found. Parameter $a$ can be expressed in closed form expression as $a=M_1 \frac{\Gamma(q)}{\Gamma(q+1/p)}$. However, $p$ and $q$ cannot be found in closed form because of the nonlinear equations arising from $M_2/M_1^2$ and $M_3/M_1^3$ \cite{Gamma}, both of which involve the parameters $p$ and $q$. These can be calculated by using least squares method or built-in functions such as \textit{fsolve} in MATLAB\cite{Gamma}. After obtaining these necessary parameters and using \eqref{eq:fy}, moment generating function (MGF) of $Y$ is expressed as

\begin{equation}
    \mathcal{M}_Y(-s)=\frac{1}{\Gamma(q)} \int_0^\infty e^{-sau^{1/p}} e^{-u} u^{q-1} du,
\end{equation}with variable change of $u=(y/a)^p$. Substituting the Mellin-Barnes integral representation of $e^{-sau^{1/p}}$ here, we get
\begin{equation}
    \mathcal{M}_Y(-s)=\frac{1}{\Gamma(q) 2\pi j} \int_\mathcal{L}\Gamma(r)(sa)^{-r}\int_0^\infty u^{q-1-r/p} e^{-u} du \,dr.
\end{equation}By employing \cite[eq.~(3.326.2)]{gradshteyn2007table}, inner integral reduces to $\Gamma(q-r/p)$ and the resulting integral can be expressed as

\begin{equation}
    \mathcal{M}_{\gamma_1}(-s)=\mathcal{M}_Y(-s)=\frac{1}{\Gamma(q)}H_{1,1}^{1,1}
\left[
a s
\;\middle|\;
\begin{matrix}
(1-q,1/p)\\
(0,1)
\end{matrix}
\right]
\label{eq:fox1}
\end{equation} using the Fox-H function definition in \cite{foxH}. After completing the MGF derivation of SNR at the first antenna, we continue by analyzing the remaining antennas. It is clear from \eqref{eq:Xk} that, $X_k= \sum_{i=1}^{N}\,|h_{ik}|\,e^{j\Delta_i}, \Delta_i=\theta_{ik}-\theta_{i1}$. Here, $\theta_{ik}$ are phases of i.i.d. Rician channels as defined before and uniformly distributed on the interval $[0,2\pi)$\cite{faz2}. Although the ordinary difference of two independent uniform phase variables follows a triangular distribution, the corresponding phase difference modulo $2\pi$ is uniformly distributed over $[0,2\pi)$. Hence, the exponential term behaves as a phasor with a uniformly distributed phase. For notational simplicity, let $Z=Z_{ik}=|h_{ik}|e^{j\Delta_i}$ and $r=|h_{ik}|$. Then, by representing $Z$ in terms of its real and imaginary components,
its two-dimensional characteristic function can be expressed as
\begin{equation}
\begin{aligned}
\Phi_Z(t_1,t_2)
&= \mathbb{E}_{r}\left[
\mathbb{E}_{\Delta_i}\left[
e^{jr\left(t_1\cos(\Delta_i)+t_2\sin(\Delta_i)\right)}
\right]\right] \\
&= \mathbb{E}_{r}\left[
J_0\left(r\sqrt{t_1^2+t_2^2}\right)
\right].
\end{aligned}
\label{eq:cf_Z}
\end{equation}
Defining $t=\sqrt{t_1^2+t_2^2}$, this reduces to
$\Phi_Z(t)=\mathbb{E}_{r}[J_0(rt)]$, as derived in Appendix~\ref{app:der}. Then, using well known PDF expression of Rician distribution, $\Phi_{Z}(t)=\int_{0}^{\infty}J_0(rt)f_{r}(r)dr$ can be expressed as
\begin{equation}\label{eq:karakter}
    \Phi_{Z}(t)=2be^{-K} \int_0^\infty r J_0(rt)e^{-br^2}I_0(2r\sqrt{Kb})\, dr,
\end{equation} where $K$ is Rician parameter, $b=\frac{1+K}{\Omega}$, $J_0(\cdot)$ and $I_0(\cdot)$ denote the zeroth-order Bessel function of the first kind and the zeroth-order modified Bessel function of the first kind, respectively \cite{gradshteyn2007table}. By making use of \cite[Eq.~(6.633.4)]{gradshteyn2007table} here, \eqref{eq:karakter} reduces to $\Phi_{Z}(t)=e^{-t^2/4b}J_0\bigg(t\sqrt{\frac{K}{b}}\bigg)$. Since $X_k$ is the sum of i.i.d. terms $Z=Z_{ik}$, the CHF of $X_k$ can be written as
\begin{equation}
    \Phi_{X_k}(t)=\Phi_{Z}(t)^N=e^{-Nt^2/4b}J_0\bigg(t\sqrt{\frac{K}{b}}\bigg)^N.
\end{equation}For analytical tractability and based on the small-argument expansion of $J_0(\cdot)$ \cite[Eq. 10.2.2]{dlmf}, we approximate $J_0\!\left(t\sqrt{\frac{K}{b}}\right) \approx \exp\left(-\frac{t^2K}{4b}\right)$, since both functions share the same expansion up to the second-order term around the origin. So, CHF of $X_k$ is given by $\Phi_{X_k}(t)=e^{-t^2(\frac{N(K+1)}{4b})}=e^{-t^2\frac{N\Omega}{4}}$, since $b=\frac{1+K}{\Omega}$. We now proceed with the CHF of $|X_k|^2=\operatorname{Re}\{X_k\}^2+\operatorname{Im}\{X_k\}^2=X_{\mathrm{Re}}^2+X_{\mathrm{Im}}^2$. So far, we have used the compact notation $t=\sqrt{t_1^2+t_2^2}$ in the argument of the CHF. Rewriting the CHF in terms of its original variables $t_1$ and $t_2$, we obtain
\begin{equation}
    \Phi_{X_k}(t)=\Phi_{X_{\mathrm{Re}},X_\mathrm{Im}}(t_1,t_2)=e^{-t_1^2\frac{N\Omega}{4}}e^{-t_2^2\frac{N\Omega}{4}}.
\end{equation}

At this point, the statistics of $X_{\mathrm{Re}}^2$ and $X_{\mathrm{Im}}^2$ must be obtained. Since the two variables are i.i.d., their PDFs can be found as $f_{X_{\mathrm{Re}}^2}(x)=f_{X_{\mathrm{Im}}^2}(x)=\frac{1}{\sqrt{\pi N\Omega x}}\exp\left(-\frac{x}{N\Omega}\right)$, $x>0$, using transformation of random variables. Then, the MGF of their sum $|X_k|^2=X_{\mathrm{Re}}^2+X_{\mathrm{Im}}^2$ is obtained as the product of their individual MGFs. The derivation is straightforward and is omitted for brevity. Finally, the MGF of $|X_k|^2$ can be found as
\begin{equation}
    \mathcal{M}_{|X_k|^2}(-s)=\frac{1}{1+N\Omega s}.
\end{equation}Hereafter, MGF of $\gamma_k,\,\,\, k=2,\cdots,N_r$ is given by
\begin{equation}\label{eq:birMGF}
    \mathcal{M}_{\gamma_k}(-s)=\mathcal{M}_{|X_k|^2}(-s\,\zeta\,\bar{\gamma})=\frac{1}{1+\zeta\, \bar{\gamma}\, N\Omega s}.
\end{equation}Every receive antenna except the reference antenna has the same statistics. Thus, by letting $\gamma_{\mathrm{Rem}}=\sum_{k=2}^{N_r}\gamma_k$ denote the SNR sum of all antennas except the reference antenna, MGF of $\gamma_{\mathrm{Rem}}$ is expressed as
\begin{equation}
\begin{aligned}
    \mathcal{M}_{\gamma_{\mathrm{Rem}}}(-s)
    &=\big(1+\underbrace{ \zeta\, \bar{\gamma}\, N\Omega }_{C}s\big)^{N_r-1} \\
    &=\frac{1}{\Gamma(N_r-1)}H_{1,1}^{1,1}
\left[
C s
\;\middle|\;
\begin{matrix}
(2-N_r,1)\\
(0,1)
\end{matrix}
\right].
\end{aligned}
\label{eq:fox2}
\end{equation}Applying \cite[Eq.~(A.13)]{foxKitap} to the product of \eqref{eq:fox1} and \eqref{eq:fox2}, the MGF of the total MRC SNR is obtained as \eqref{eq:MGF}, shown at the top of the this page. Applying $\mathcal{L}^{-1}\big\{\mathcal{M}_{\gamma_\mathrm{MRC}}(-s)/s\big\}$, the CDF of the total MRC SNR, $F_{\gamma_{\mathrm{MRC}}}(\cdot)$ is directly derived as \eqref{eq:CDF}, at the top of this page. It should be noted that these expressions in the form of Fox-H function follow the definition in \cite{foxH} and can be efficiently evaluated using the Python implementation provided therein.

\begin{figure*}[!t]
\centering
\vspace{-0.95em}
\begin{equation}
\scalebox{1.1}{$
\mathcal{M}_{\gamma_{\mathrm{MRC}}}(-s)=
\frac{1}{\Gamma(q)\Gamma(N_r-1)}
H^{\,0,0:\,1,1;\,1,1}_{\,0,0:\,1,1;\,1,1}
\!\left[
\begin{array}{c}
as \\[-2pt]
Cs
\end{array}
\;\middle|\;
\begin{array}{l}
-
\,\,: (1-q,1/p)\ ;\ (2-N_r,1)\\[3pt]
-\ :\,\,\,\,\,\, (0,1)\qquad ;\ \,\,\,\,\,\,(0,1)
\end{array}
\right]
$}
\label{eq:MGF}
\end{equation}
\vspace{-0.25em}
\end{figure*}

\begin{figure*}[!t]
\centering
\vspace{-0.95em}
\begin{equation}
\scalebox{1.1}{$
F_{\gamma_{\mathrm{MRC}}}(x)=
\frac{1}{\Gamma(q)\Gamma(N_r-1)}
H^{\,0,0:\,1,1;\,1,1}_{\,1,0:\,1,1;\,1,1}
\!\left[
\begin{array}{c}
a/x \\[-2pt]
C/x
\end{array}
\;\middle|\;
\begin{array}{l}
(1:1,1)
\,\,: (1-q,1/p)\ ;\ (2-N_r,1)\\[3pt]
\qquad -\,\,\,\,\,\ :\,\,\,\,\,\, (0,1)\qquad ;\ \,\,\,\,\,\,(0,1)
\end{array}
\right]
$}
\label{eq:CDF}
\end{equation}
\vspace{-0.25em}
\end{figure*}

\subsection{RAS Analysis}

We continue by analyzing the system model using the RAS technique. In this technique, after the training period, RF chain switches to the receive antenna yielding maximum SNR. Unlike the MRC case, only one receive antenna is active under RAS. Therefore, the phase-shifts of RIS can be fully aligned with the channels corresponding to selected receive antenna. Hence, for any $k=1,\ldots,N_r$, the total channel in \eqref{eq:ilk} can be written as $X_k=\sum_{i=1}^{N}|h_{ik}|$ and \eqref{eq:fox1} applies to the SNR at each receive antenna. Employing $F_{\gamma_k}(x)=\mathcal{L}^{-1}\big\{\mathcal{M}_{\gamma_k}(-s)/s \big\}$ in \eqref{eq:fox1}, CDF of SNR at each antenna can be directly derived as
\begin{equation}
    F_{\gamma_k}(x)=\frac{1}{\Gamma(q)}H_{2,1}^{1,1}
\left[
\frac{a}{x}
\;\middle|\;
\begin{matrix}
(1-q,1/p),(1,1)\\
(0,1)
\end{matrix}
\right].
\label{eq:foxRAS}
\end{equation}With the aid of higher order statistics \cite{benimki}, total RAS SNR can be written as $F_{\gamma_{\mathrm{RAS}}}(x)=\big[F_{\gamma_k}(x)\big]^{N_r}$. Applying \cite[Eq.~(A.13)]{foxKitap} and using the notation in \cite{foxH}, CDF of total SNR using RAS is given in the form of $N_r$-variate Fox-H function by \eqref{eq:RASCDF} at the top of this page, where $\xi_1=\xi_2=\cdots=\xi_{N_r}=a/x$. Note that, average SNR actually multiplies parameter $a$ since $M_1=\zeta \bar{\gamma} \mathbb{E}[S_N^{\,2}]$. Therefore, $\bar{\gamma}$ is existent within (\ref{eq:MGF}) to (\ref{eq:RASCDF}). Also OP for both systems are obtained by $P_{\mathrm{out}}=F_{\gamma_{\mathrm{MRC}}}(\gamma_{\mathrm{th}})$ and $P_{\mathrm{out}}=F_{\gamma_{\mathrm{RAS}}}(\gamma_{\mathrm{th}})$ for MRC and RAS, respectively.

\begin{figure*}[!t]
\centering
\vspace{-0.95em}
\begin{equation}
\scalebox{1.1}{$
F_{\gamma_{\mathrm{RAS}}}(x)=
\frac{1}{\big[\Gamma(q)\big]^{N_r}}
H^{\,0,0:\,[1,1]_{1:N_r}}_{\,0,0:\,[2,1]_{1:N_r}}
\!\left[
\begin{array}{c}
\xi_1 \\[-2pt]
\vdots\\[-2pt]
\xi_{N_r}
\end{array}
\;\middle|\;
\begin{array}{l}
-\,:\,
\left[
\begin{array}{c}
(1-q,1/p)\ (1,1)\\[3pt]
(0,1)
\end{array}
\right]_{1:N_r}
\end{array}
\right]
$}
\label{eq:RASCDF}
\end{equation}
\vspace{-0.25em}
\end{figure*}

\section{Asymptotic Behavior}
In asymptotic analysis, it is critical to define the starting point. Since the OP expressions of MRC and RAS are based on the $\Gamma\Gamma$ approximation parameters, they approximate the exact behavior of the system very well over a significant SNR range. However, as $\bar{\gamma}\to\infty$, DO of $\Gamma\Gamma$ distribution does not yield DO of exact distribution. So, the original PDF expressions should be used to accurately characterize the asymptotic behavior. Also, note that as $\bar{\gamma}\to\infty$, the arguments of both \eqref{eq:CDF} and \eqref{eq:RASCDF} tend to zero. Therefore, the asymptotic analysis must be carried out using the small-argument approximations of the channel distribution.
\subsection{Asymptotic RAS Analysis}
Since all channels are compensated in terms of phase, SNR of $k$th antenna is $\gamma_k=\zeta\bar{\gamma}X_1^2$. We begin the analysis by considering single antenna by denoting $\gamma_1=\zeta\bar{\gamma}X_1^2=\zeta\bar{\gamma} W$. Here, denoting $|h_{i1}|=\rho_{i}$, $W$ can be written as 
\begin{equation}\label{eq:19denk}
    W=\underbrace{\sum_{i=1}^{N}\rho_{i}^2}_{Q}+2\sum_{i>j}\rho_{i}\rho_{j}\,,
\end{equation}where $Q$ is introduced for clear notation. Although the statistics of $Q$ are analytically tractable, the cross-term part does not readily lead to a tractable and insightful analysis. We show in Appendix~\ref{app:sonder} that, term $Q$ is sufficient to capture the asymptotic behavior of $W$. In other words, $W$ and $Q$ yield the same DO. Hence, the asymptotic behavior of $\gamma_1$, which is obtained by scaling $W$, is directly determined by DO of $Q$. However, as $K$ increases, the term $Q$ alone does not provide an accurate approximation over the observable SNR region. To achieve a closer approximation, we denote 
\begin{equation}
    W=Q\frac{W}{Q}=Q\frac{\big(\sum_{i=1}^{N}\rho_{i}\big)^2}{\sum_{i=1}^{N}\rho_{i}^2}=QG \approx QG_{\mathrm{eff}}.
\end{equation}Since $Q$ captures the DO of $W$, the remaining term is incorporated through its expectation to obtain a more accurate yet still tractable approximation. Here, $G_{\mathrm{eff}}=\frac{\mathbb{E}[W]}{\mathbb{E}[Q]}$, $\mathbb{E}[Q]=\sum_{i=1}^{N}\mathbb{E}[\rho_{i}^2]=N\Omega$ and $\mathbb{E}[W]=\sum_{i=1}^{N}\mathbb{E}[\rho_{i}^2]+2\sum_{i>j}\mathbb{E}[\rho_{i}]\mathbb{E}[\rho_{j}]=N\Omega+N(N-1)m_1^2$. Note that, $m_1$ was denoted as first moment of single Rician variable. Consequently, $G_{\mathrm{eff}}= 1 + (N-1)\frac{m_1^2}{\Omega}$ can be found. We now proceed to distribution of $Q$. Since $Q$ is sum of squared Rician variables, it is known from \cite{chi2} that, $Q\sim\frac{\Omega}{2(K+1)}\chi^2_{2N}(2NK)$. Here, $\chi^2_{\nu}(\lambda)$ denotes the noncentral chi-square distribution with degrees of freedom and noncentrality parameters $\nu$ and $\lambda$, respectively. Re-expressing $\gamma_1=\zeta\bar{\gamma}\,G_{\mathrm{eff}}\frac{\Omega}{2(K+1)}\chi^2_{2N}(2NK)=C_0 Q_0$ where $C_0=\zeta\bar{\gamma}\,G_{\mathrm{eff}}\frac{\Omega}{2(K+1)}$ and $Q_0 = \chi^2_{2N}(2NK)$, asymptotic MGF of $\gamma_1$ can be written as 
\begin{equation}
\begin{aligned}
\mathcal{M}_{\gamma_1}^{\infty}(-s)
&= \mathbb{E}[e^{-s\gamma_1}]
 = \mathbb{E}[e^{-sC_0 Q_0}]
 = \mathcal{M}_{Q_0}^{\infty}(-sC_0) \\
&= (1+2C_0s)^{-N} \exp\bigg(-\frac{2NKC_0s}{1+2C_0s}\bigg),
\end{aligned}
\end{equation}where the Laplace transform of $Q_0$ is given by a well-known closed-form expression. Here, $C_0\to\infty$ as $\gamma_1\to\infty$. So, we approximate $(1+2\,C_0s)^{-N}\approx(2\,C_0s)^{-N}$. By substituting $\exp\left(-\frac{2NKC_0s}{1+2C_0s}\right)=\exp(-NK)\exp\left(\frac{NK}{1+2C_0s}\right)$ and applying the series expansion of the latter exponential term, we obtain
\begin{equation}\label{eq:222}
    \mathcal{M}_{\gamma_1}^{\infty}(-s)=\frac{e^{-NK}}{(2C_0)^N}\sum_{k=0}^{\infty}\frac{(NK)^k}{k!}s^{-N}(1+2C_0s)^{-k}.
\end{equation}Asymptotic CDF can be obtained via $\mathcal{L}^{-1}\big\{\mathcal{M}_{\gamma_1}^{\infty}(-s)/s\big\}$. Here, the inverse Laplace transforms of the two $s$-dependent terms are taken separately, and the final result is obtained through their convolution. The two transforms are given as
\begin{equation}
\begin{aligned}
f_1(x)&=\mathcal{L}^{-1}\big\{ s^{-N-1} \big\}
= \frac{x^N}{\Gamma(N+1)}
, \\
f_2(x)&=\mathcal{L}^{-1}\big\{ (2C_0)^{-k}(1/2C_0+s)^k \big\}
= \frac{x^{k-1}e^{-x/2C_0}}{(2C_0)^k\Gamma(k)}.
\end{aligned}
\end{equation}Since the derivations of $f_1(x)$ and $f_2(x)$ are straightforward, they are omitted for brevity. Let
$F(x)=\int_{0}^{x}f_1(t)f_2(x-t)\,dt
=\frac{1}{\Gamma(N+1)(2C_0)^k\Gamma(k)}
\int_{0}^{x}t^N(x-t)^{k-1}e^{-(x-t)/(2C_0)}\,dt$.
Then, the changes of variables $t=xu$ and $v=1-u$ are applied successively. Finally, employing \cite[Eq. 13.4.1]{dlmf} in the resulting integral yields
\begin{equation}\label{eq:fxx}
    F(x)=\frac{x^{N+k}}{(2C_0)^k\Gamma(N+k+1)}{}_1F_1\bigg(k;N+k+1;-\frac{x}{2C_0}\bigg),
\end{equation}where ${}_1F_1(\cdot)$ denotes the confluent hypergeometric function \cite{gradshteyn2007table}. Using this inverse Laplace transform and \eqref{eq:222}, we obtain
\begin{equation}\label{eq:asymTEK}
\begin{aligned}
F_{\gamma_1}^{\infty}(x)
&=\frac{e^{-NK}}{(2C_0)^N}
\sum_{k=0}^{\infty}
\frac{(NK)^kx^{N+k}}{k!(2C_0)^k\Gamma(N+k+1)} \\
&\quad \times
{}_1F_1\bigg(k;N+k+1;-\frac{x}{2C_0}\bigg).
\end{aligned}
\end{equation}Asymptotic OP of single antenna is written by $P_{\mathrm{out}}^{\infty}=F_{\gamma_1}^{\infty}(\gamma_{\mathrm{th}})$. Here, considering the DO definition in \cite{DO24}, the first term of the summation determines the asymptotic behavior. Since the remaining terms contain higher-order inverse powers of $\bar{\gamma}$, they vanish under this definition as $\bar{\gamma}\to\infty$. Similarly, the higher-order terms in the series expansion of ${}_1F_1(\cdot)$ also go to zero as $\bar{\gamma}\to\infty$. Since $C_0=\zeta\bar{\gamma}G_{\mathrm{eff}}\frac{\Omega}{2(K+1)}$ is proportional to $\bar{\gamma}$, the DO can be obtained as
\begin{equation}
    \mathrm{DO}=\lim_{\bar{\gamma} \to \infty} -\frac{\log(P_{\mathrm{out}}^{\infty})}{\log(\bar{\gamma})}=N.
\end{equation}
Note that, this result was obtained considering single antenna. Asymptotic OP of the system using RAS can be expressed by having $N_r$th power of \eqref{eq:asymTEK}, i.e., $F_{\gamma_{\mathrm{RAS}}}^{\infty}(\gamma_{\mathrm{th}})=\big[F_{\gamma_{1}}^{\infty}(\gamma_{\mathrm{th}})\big]^{N_r}$, which multiplies the asymptotic exponent by $N_r$. Then, DO of RAS system can directly be found as $\mathrm{DO}_{\mathrm{RAS}}=NN_r$ \cite{benimki}.

\subsection{Asymptotic MRC Analysis}
In line with \eqref{eq:Xk}, only one receive antenna has phase-free statistics, while the remaining antennas experience uniformly distributed residual phases. The asymptotic MGF of the SNR at the phase-free antenna was derived in \eqref{eq:222}. For the remaining antennas, the exact MGF expression was derived in \eqref{eq:fox2}. Hence, this expression can be directly employed in the asymptotic analysis. Here, approximated asymptotic MGF is expressed as
\begin{equation}\label{eq:rgb}
    \mathcal{M}_{\gamma_{\mathrm{Rem}}}^{\infty}(-s)=(1+\zeta\, \bar{\gamma}\, N\Omega s)^{-(N_r-1)}\, \overset{\bar{\gamma}\to\infty}{\approx}(\underbrace{\zeta\, \bar{\gamma}\, N\Omega}_{D} s)^{-(N_r-1)}.
\end{equation}where $D$ is defined for notational convenience. So, using \eqref{eq:222} together with \eqref{eq:rgb}, asymptotic MGF of MRC SNR can be expressed as
\begin{equation}
\begin{aligned}
\mathcal{M}_{\gamma_{\mathrm{MRC}}}^{\infty}(-s)
&= \mathcal{M}_{\gamma_1}^{\infty}(-s)\mathcal{M}_{\gamma_{\mathrm{Rem}}}^{\infty}(-s)\\
&\hspace{-5em}=\frac{e^{-NK}D^{-(N_r-1)}}{(2C_0)^N}
\sum_{k=0}^{\infty}\frac{(NK)^k}{k!}
s^{-(N+N_r-1)}(1+2C_0s)^{-k}.
\end{aligned}
\end{equation}The asymptotic CDF can be obtained by applying
$\mathcal{L}^{-1}\big\{\mathcal{M}_{\gamma_{\mathrm{MRC}}}^{\infty}(-s)/s\big\}$
and proceeding as in the derivation of \eqref{eq:fxx}. Consequently, asymptotic CDF of MRC SNR can be derived as
\begin{equation}
\begin{aligned}
        F_{\gamma_{\mathrm{MRC}}}^{\infty}(x)&=\frac{e^{-NK}}{D^{-(N_r-1)}(2C_0)^N}\sum_{k=0}^{\infty}
\frac{(NK)^k x^{N+N_r-1+k}}{k!(2C_0)^k\Gamma(N+N_r+k)} \\
&\quad \times
{}_1F_1\bigg(k;N+N_r+k;-\frac{x}{2C_0}\bigg).
\end{aligned}
\end{equation}By following the same reasoning as in the previous subsection, only the first term of the summation contributes to the asymptotic behavior under the DO definition in \cite{DO24}. The remaining terms, including the higher-order terms of ${}_1F_1(\cdot)$, vanish as $\bar{\gamma}\to\infty$. Hence, the DO of the MRC system can be derived similarly to the previous subsection. Note that $D=\zeta\bar{\gamma}N\Omega$ and 
$C_0=\zeta\bar{\gamma}G_{\mathrm{eff}}\frac{\Omega}{2(K+1)}$, both of which are proportional to $\bar{\gamma}$. Substituting the asymptotic OP expression ($P_{\mathrm{out}}^{\infty}=F_{\gamma_{\mathrm{MRC}}}^{\infty}(\gamma_{\mathrm{th}})$) into the DO definition \cite{DO24}, we obtain
\begin{equation}
\mathrm{DO_{MRC}}=\lim_{\bar{\gamma} \to \infty} -\frac{\log(P_{\mathrm{out}}^{\infty})}{\log(\bar{\gamma})}=N+N_r-1.
\end{equation}This is an interesting result, as the phase-free antenna contributes $N$ to the total DO, whereas each remaining antenna increases the DO only by one. Thus, the total channel observed at each remaining receive antenna behaves as an effective single-antenna channel due to the uniformly distributed residual phase component.

\section{Numerical Results}
In this section, we present the theoretical curves together with Monte Carlo simulation results. The SNR threshold is set to $\gamma_{\mathrm{th}}=1$, which corresponds to a target rate of $R_b=1$ bit/s/Hz through $\gamma_{\mathrm{th}}=2^{R_b}-1$ for all of the results. Path loss coefficient is expressed as $\zeta=d^{-\alpha}$, where $d$ and $\alpha$ denote distance (in meters) and the path loss exponent, respectively. Note that we use $F_{\gamma_{\mathrm{RAS}}}(\gamma_{\mathrm{th}})$ with $N_r=1$ for the single input single output (SISO) results, since the MRC expressions are derived considering multiple antennas, whereas the RAS expressions also cover the single-antenna scenario. Unless otherwise stated, the Rician factor is set to $K=1$. In the figures, markers show simulation results.

\begin{figure}[t]
  \centering
  \includegraphics[width=0.9\columnwidth]{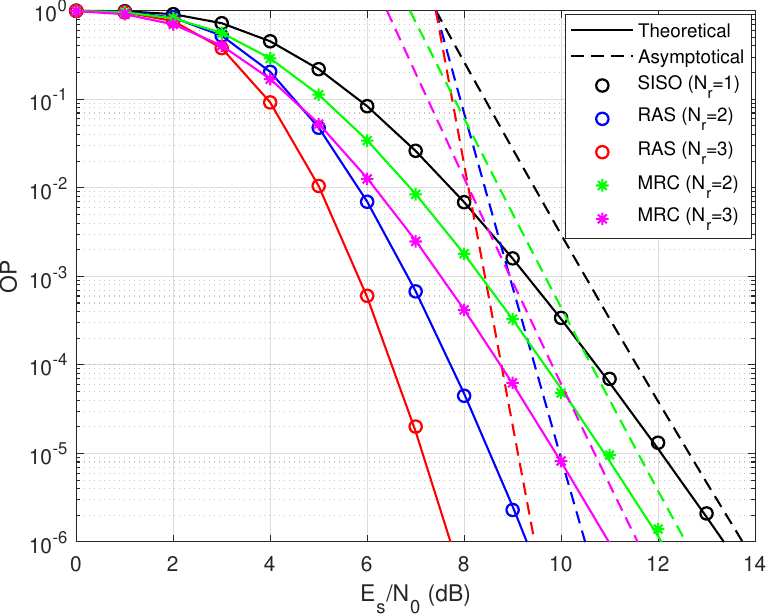}
  \caption{OP comparison of RAS and MRC systems with the SISO benchmark for $N=8$.}
  \label{fig:ex1}
\end{figure}

In Fig.~\ref{fig:ex1}, MRC and RAS results including asymptotical ones are given with having SISO as a baseline and using $d=10$, $\alpha=2.1$, corresponding to a relatively favorable propagation environment with mild path loss attenuation. It can be observed that using RAS with adding one and two antennas yields about $3.58$ dB and $4.94$ dB SNR gains at an OP level of $10^{-5}$ compared to SISO system, respectively. On the other hand, using MRC with one and two additional antennas yields only $1.15$ dB and $2.1$ dB SNR gains at the same OP level, respectively. Clearly, due to the practical phase-shift constraint of RISs, RAS outperforms MRC considerably for the same number of antennas. It is worthy to note that, for relatively small numbers of RIS elements, the DO can be inferred from the slopes of the asymptotic curves. It is clear that RAS curves have higher DO than the same system configuration using MRC, in line with the theoretical derivations.

\begin{figure}[t]
  \centering
  \includegraphics[width=0.9\columnwidth]{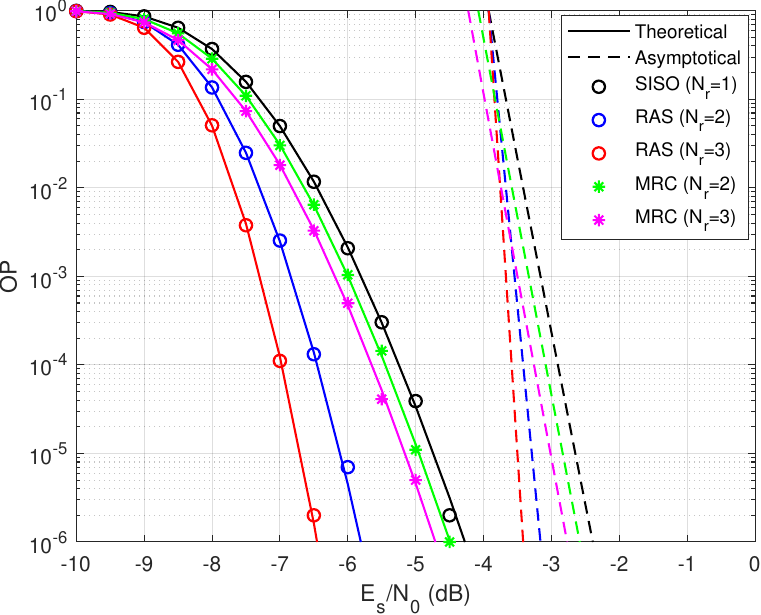}
  \caption{OP comparison of RAS and MRC systems with the SISO benchmark for $N=32$.}
  \label{fig:ex2}
\end{figure}

Moving on to Fig.~\ref{fig:ex2}, we use the same configuration used in Fig.~\ref{fig:ex1}, by changing only the number of RIS elements to $N=32$. For RAS, increasing the number of receive antennas by one and two provides SNR gains of $1.4$ dB and $1.98$ dB at an OP of $10^{-5}$, respectively. Whereas, when employing MRC, the same antenna additions only provide $0.2$ dB and $0.42$ dB SNR gains compared to SISO case at the same OP level. Taking both Fig.~\ref{fig:ex1} and Fig.~\ref{fig:ex2} into account, it can be inferred that, RAS provides considerably better performance than MRC across different numbers of RIS elements. Although the absolute SNR gain provided by additional receive antennas in RAS decreases as the number of RIS elements increases, the relative performance advantage of RAS over MRC becomes more pronounced.

\begin{figure*}[!t]
  \centering
  \includegraphics[width=0.9\textwidth]{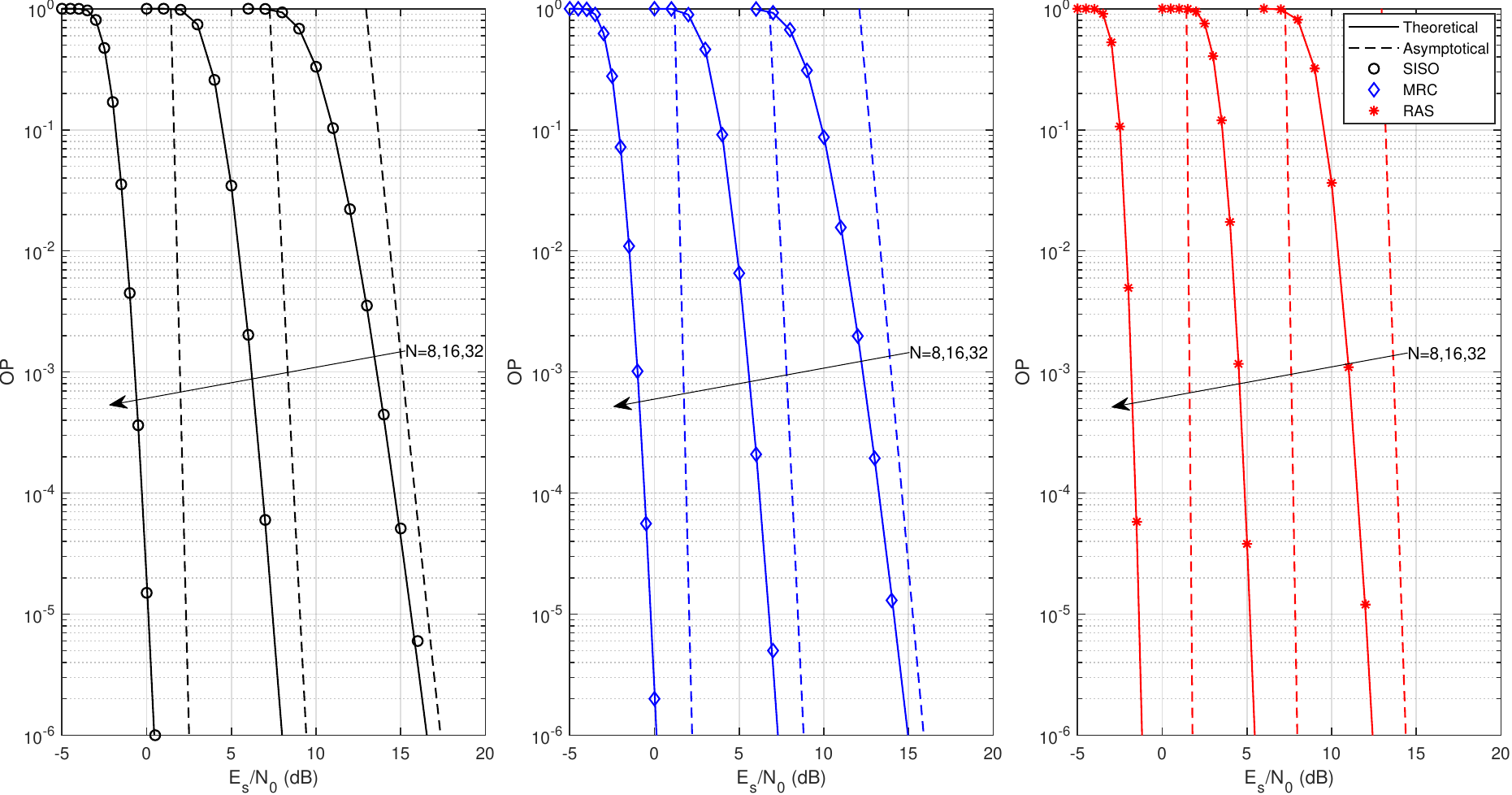}
  \caption{OP performance comparison of SISO, RAS, and MRC systems for different numbers of RIS elements, i.e., $N=8,16,32$.}
  \label{fig:ex3}
\end{figure*}

Fig.~\ref{fig:ex3} depicts the effect of increasing the number of RIS elements among multiple and single antenna cases. Number of receive antennas for RAS and MRC configurations is set to $N_r=3$. The distance and path loss exponent are chosen as $d=15$ and $\alpha=2.3$, respectively, which corresponds to a lightly obstructed line-of-sight (LoS)-dominant propagation environment. Accordingly, a moderate Rician factor is considered by setting $K=3$. At an OP of $10^{-5}$, increasing the number of RIS elements from $8$ to $16$ yields SNR gains of $6.89$ dB, $7.47$ dB, and $8.16$ dB for RAS, MRC, and SISO, respectively. Similarly, increasing the number of RIS elements from $16$ to $32$ yields SNR gains of $6.43$ dB, $6.83$ dB, and $7.35$ dB, respectively. It is observed that increasing the number of RIS elements improves the OP performance of all schemes. The SNR gain is slightly more pronounced for SISO and MRC, since these systems have a larger improvement margin compared with RAS. However, RAS already operates in a better performance region and consistently outperforms MRC over the considered RIS element configurations.

\section{Conclusion}
In this work, we have examined an overlooked issue in the literature on RIS-assisted MRC systems, namely the \textit{phase dependency issue}. Performance analyses of RIS-assisted MRC and RAS systems have been carried out to better emphasize the performance degradation in MRC considering practical limitations. Closed form OP expressions in the form of Fox-H function have been derived and asymptotic analyses have been carried out to show that, in the proposed systems, the DOs of RAS and MRC are given by $N N_r$ and $N+N_r-1$, respectively. Furthermore, we show that RAS provides considerably better performance than MRC across different numbers of RIS elements in the proposed systems. Together with its simple single-RF-chain receiver structure, these results highlight the potential of RAS for hardware-efficient RIS-empowered transmitter architectures in next-generation IoT networks.

\appendices
\section{}
\label{app:ilkder}
We define $S_N=\sum_{i=1}^{N}|h_{i1}|=\sum_{i=1}^{N}\rho_{i}$. $m_r=\mathbb{E}[\rho_{i}^r]$ denotes the $r$th moment of the Rician random variable, which has a well-known closed-form expression. Proceeding with the definitions $S_N=S_{N-1}+\rho_{N}$ and
\[
S_N^r=(S_{N-1}+\rho_{N})^r
=\sum_{j=0}^{r}\binom{r}{j}S_{N-1}^{r-j}\rho_{N}^j,
\]
the expectation of $S_N^r$ can be expressed as
\begin{equation}
    \mathbb{E}[S_N^r]=\sum_{j=0}^{r}\binom{r}{j}\mathbb{E}[S_{N-1}^{r-j}]\,m_j,
\end{equation}with the convention $m_0=1$. 
\section{}
\label{app:der}
Consider a complex variable $Z=R\,e^{j\varphi}=A+jB$. Here, $A=R\cos(\varphi),B=R\sin(\varphi)$. Let $R$ be an arbitrary random variable and $\varphi$ a phase uniformly distributed over $[0,2\pi]$. Then, joint CHF can be written as
\begin{equation}
    \Phi_{A,B}(t_1,t_2)=\mathbb{E}[e^{j(t_1A+t_2B)}]=\mathbb{E}[e^{jR(t_1\cos(\varphi)+t_2\sin(\varphi))}].
\end{equation}Latter expression is given by
\begin{equation}
\begin{aligned}
&\mathbb{E}_{R,\varphi}
\left[
e^{jR\big\{ t_1\cos(\varphi)+t_2\sin(\varphi) \big\}}
\right] \\
&\qquad =
\mathbb{E}_{R}
\left[
\frac{1}{2\pi}
\int_{0}^{2\pi}
e^{jR\big\{t_1\cos(\varphi)+t_2\sin(\varphi)\big\}}
d\varphi
\right].
\end{aligned}
\end{equation}Here, with variable changes of $t_1=t\cos(\alpha),t_2=t\sin(\alpha)$ and $t=\sqrt{t_1^2+t_2^2}$, integral reduces to $J_0(Rt)$. Finally, CHF is expressed as
\begin{equation}
    \Phi_{Z}(t)=\Phi_{A,B}(t_1,t_2)=\mathbb{E}_R\bigg[J_0\bigg(R\sqrt{t_1^2+t_2^2}\bigg)\bigg]=\mathbb{E}_R[J_0(Rt)].
\end{equation}

\section{}
\label{app:sonder}
Consider CDF of SNR at single antenna as $P(\gamma_1\leq\gamma_{\mathrm{th}})=P(\zeta\bar{\gamma}\,W\leq\gamma_{\mathrm{th}})=P\big(W\leq\frac{\gamma_{\mathrm{th}}}{\zeta\bar{\gamma}}\big)=F_W\big(\frac{\gamma_{\mathrm{th}}}{\zeta\bar{\gamma}}\big)$. Here, $W$ is defined as in \eqref{eq:19denk} and $P(\alpha)$ is probability of $\alpha$. In the following analysis, the small-argument regime $w\to0$ corresponds to the high-SNR regime $\bar{\gamma}\to\infty$, since the outage threshold associated with $W$ is inversely proportional to $\bar{\gamma}$. Having that said, for the phase-free antenna, we have
\[
W=\left(\sum_{i=1}^{N}|h_{i1}|\right)^2
=\left(\sum_{i=1}^{N}\rho_{i}\right)^2 .
\]
Using \eqref{eq:19denk}, the lower and upper bounds of $W$ are respectively defined as
\begin{equation}\label{eq:35}
W
=\sum_{i=1}^{N}\rho_{i}^2+2\sum_{i>j}\rho_{i} \rho_{j}
\geq \sum_{i=1}^{N}\rho_{i}^2=Q.
\end{equation}

\begin{equation}\label{eq:36}
\begin{aligned}
W
=\left(\sum_{i=1}^{N}1\cdot \rho_{i}\right)^2 &\leq \left(\sum_{i=1}^{N}1^2\right)
\left(\sum_{i=1}^{N}\rho_{i}^2\right) \\
&=N\left(\sum_{i=1}^{N}\rho_{i}^2\right)
=NQ.
\end{aligned}
\end{equation}Here, \eqref{eq:35} arises from the fact that all channel amplitudes are positive and \eqref{eq:36} arises from Cauchy-Schwarz inequality. Thus, the lower and upper bounds of $Q$ can be summarized as
\begin{equation}\label{eq:37denk}
    \frac{W}{N}\leq Q\leq W.
\end{equation}Here, we derive bounds on the CDF of $W$ in order to investigate its asymptotic behavior. Let us write $W=Q+T$, where $T$ represents the cross-term component in \eqref{eq:19denk}. Then, $F_W(w)$ can be written as
\begin{equation}
    P(Q+T\leq w) = \int_{0}^{\infty}P(Q \leq w-t \mid T=t)f_T(t) dt.
\end{equation}It is evident that $T$ is a non-negative random variable. Since the conditional CDF is a non-decreasing function, the following inequality holds
\begin{equation}
\begin{aligned}
&\int_{0}^{\infty}
P(Q \leq w-t \mid T=t) f_T(t)\,dt \\
&\qquad \leq
\int_{0}^{\infty}
P(Q \leq w \mid T=t) f_T(t)\,dt ,
\end{aligned}
\end{equation}which directly corresponds to
\begin{equation}
    F_W(w) \leq F_Q(w).
\end{equation}Thus, we have $Q \leq_{\mathrm{st}} W$, i.e., $Q$ is smaller than $W$ in the usual stochastic order \cite[Sec.~1.A]{stoch}. We also know from \eqref{eq:37denk} that, $W \leq QN$. Similarly, it can be shown that $W \leq_{\mathrm{st}} QN$. Finally, CDF of $W$ can be bounded as
\begin{equation}\label{eq:sirala}
    F_Q(w/N) \leq F_W(w) \leq F_Q(w).
\end{equation}To determine the asymptotic behavior, we examine the small-argument behavior of $W$. Since $Q$ is the sum of squared Rician random variables, it follows a noncentral chi-square distribution \cite{chi2}. Thus, the CDF of $Q$ can be expressed as $F_Q(w)=1-\boldsymbol{\mathrm{Q}}_N\big(\sqrt{2NK},\sqrt{w\,\Omega/2\,(K+1)}\big)$, where $K$, $\Omega$ and $\boldsymbol{\mathrm{Q}}_m(\cdot,\cdot)$ denote Rician factor, average channel power and the generalized Marcum-$\boldsymbol{\mathrm{Q}}$ function, respectively. With integral representation of Marcum-$\boldsymbol{\mathrm{Q}}$ function \cite[Eq.~(4.59)]{simon} and $\beta=\sqrt{w\,\Omega/2\,(K+1)})$ for convenience, CDF of $Q$ is expressed as
\begin{equation}\label{eq:appendixte}
\begin{aligned}
F_Q(w)=1-\boldsymbol{\mathrm{Q}}_N\big(\sqrt{2NK},\beta\big)
&=
\frac{1}{\big(\sqrt{2NK}\big)^{N-1}}\\
&\hspace{-12em} \times\int_{0}^{\beta}
x^N
\exp\left(-\frac{x^2+2NK}{2}\right)
I_{N-1}\big(\sqrt{2NK}\,x\big)\,dx .
\end{aligned}
\end{equation}Since the small-argument behavior of $W$ is examined, the corresponding argument $\beta$ also tends to zero. Therefore, $\exp(-x^2/2) \to1$ and using first term of series expansion of $I_{N-1}\big(\sqrt{2NK} x\big)$ \cite[Eq.~(9.6.10)]{abramowitz} for small argument behavior, \eqref{eq:appendixte} reduces to
\begin{equation}\label{eq:43}
    F_Q(w) \approx \frac{e^{-NK}2(K+1)}{2^NN!\,\Omega}w^N,
\end{equation}which can be simply expressed as a product of a coefficient and a power term as
$F_Q(w)\approx c w^N$ for notational convenience. Accordingly, the lower bound CDF is obtained as
$F_Q(w/N)\approx \frac{c}{N^N}w^N$. Since $w$ is positive, dividing each term of \eqref{eq:sirala} by $w^N$ does not alter the inequality. Then, dividing \eqref{eq:sirala} by $w^N$, we obtain
\begin{equation}\label{eq:sirala2}
    \frac{c}{N^N} \leq \frac{F_W(w)}{w^N} \leq c.
\end{equation}It is evident that, as $w \to 0$, the term in the middle is bounded by two finite values. Hence, $F_W(w)/w^N$ is also a finite number. Also, since $w \to 0$ as $\bar{\gamma} \to \infty$, $\log(\bar{\gamma}) \sim -\log(w)$ holds and DO definition from \cite{DO24} can be expressed as
\begin{equation} \label{eq:DOapp}
    \mathrm{DO} = \lim_{\bar{\gamma}\to \infty} - \frac{\log(F_W(w))}{\log(\bar{\gamma})}=\lim_{w\to 0}  \frac{\log(F_W(w))}{\log(w)}.
\end{equation}Thereafter, we rewrite $F_W(w)=\frac{F_W(w)}{w^N}w^N$. Having logarithm of this expression we get
\begin{equation}
    \log(F_W(w))=\log\bigg(\frac{F_W(w)}{w^N}\bigg) + N\log(w).
\end{equation}Substituting this into \eqref{eq:DOapp}, we get
\begin{equation}\label{eq:appSON}
    \mathrm{DO}_W=\lim_{w\to 0}\frac{\log\big(\frac{F_W(w)}{w^N}\big)}{\log(w)} + \frac{N\log(w)}{\log(w)}=N,
\end{equation}as the first term is the ratio of a finite number to $-\infty$. Here, $\mathrm{DO}_{W}$ represents the DO of $W$. Similarly, DO of $Q$ can be found as $\mathrm{DO}_Q=N$. Alternatively, $\mathrm{DO}_{W}$ can be determined using a squeeze approach based on \eqref{eq:sirala}. The DOs of $F_Q(w)\approx cw^N$ and
$F_Q(w/N)\approx \frac{c}{N^N}w^N$ are both $N$. Therefore, applying
the DO definition to \eqref{eq:sirala} yields
\begin{equation}
    N \leq \mathrm{DO}_{W} \leq N.
\end{equation}
 Since the lower and upper bounds coincide, it follows that $\mathrm{DO}_{W}=N$. Thus, as shown by both approaches, $F_W(w)$ and $F_Q(w)$, corresponding to the squared Rician sum and the sum of squared Rician random variables, respectively, have the same DO. Also note that, this approach is applicable to any channel distribution that exhibits small argument behavior similar to \eqref{eq:43}. This completes the proof.

\section*{Acknowledgment}
The work of M. F. Karabay and O. Kucur is supported in part by TÜBİTAK 1515 Frontier R\&D Laboratories Support Program for Türk Telekom 6G R\&D Lab under project number 5249902 and in part by TÜBİTAK BİLGEM HİSAR Lab.
The work of E. Basar is supported by the Research Council of Finland under Grant 380594 (FORMULA3 project).

\bibliographystyle{IEEEtran}
\bibliography{references}

\end{document}